\documentclass{kluwer}
\usepackage{psfig}

\begin{document}

\begin{article}
\begin{opening}
\title{Empirical Population Synthesis for 74 Blue Compact Galaxies
\thanks{supported by CNNSF 10073009 and the Alexander von Humboldt 
Foundation.}}
\author{X. \surname{KONG}\email{xkong@ustc.edu.cn}}
\runningauthor{X. KONG}
\runningtitle{Empirical population synthesis for BCGs}
\institute{Max Planck Institute for Astrophysics, D-85741 Garching, 
Germany\\
Center for Astrophysics, Univ. of Sci. and Tech. of China, 230026, 
P. R. China}

\begin{abstract}

We have observed the largest optical spectra sample of 74 blue 
compact galaxies. Stellar population properties of them were derived 
by comparing the equivalent widths of strong absorption features 
and continuum colors, using a method of empirical population 
synthesis based on star cluster sample.  The results indicate that 
blue compact galaxies are typically age-composite stellar system, 
the continuum flux fractions at 5870\AA\ due to old stellar 
components and young stellar components are both important for most 
of the galaxies.  The stellar populations of blue compact galaxies 
present a variety of characteristics, and the contribution from 
different age and metallicity components is different.  The star 
formation episodes are usually short, some galaxies maybe 
undergoing their first global episode of star formation, while for 
the most sample galaxies, older stars contribute to at most half 
the optical emission.  Our results suggest that BCGs are old 
galaxies, in which star formation occurs in short intense burst 
separated by long quiescent phases.

\end{abstract}

\keywords{
galaxies: compact -- galaxies: stellar content -- galaxies: star 
clusters}
\end{opening}

\section{Introduction}

Blue compact galaxies (BCGs) are characterized by their very blue 
color, compact appearance, high gas content, strong nebular 
emission lines, and low chemical abundances. Detailed studies of 
BCGs are important not only for understanding their intrinsic 
properties, but also for understanding star formation processes and 
galaxy evolution in different environments.  To resolve the stellar 
components and better constrain the star formation histories of BCGs, 
we analyze the optical spectra of 74 BCGs with a Empirical Population 
Synthesis (EPS) technique, which has been pioneered by the work of 
Bica (1988) and developed by the work of Cid Fernandes et al. (2001).

Our main goals are as follows:
1) Resolve the stellar populations of BCGs;
2) Reconstruct the star formation histories of BCGs;
3) Subtract the underlying stellar absorption spectrum from the 
observed galaxy spectrum, and to obtain the emission line spectrum.

\section{Observations}

We have observed optical spectra of 97 blue compact galaxies with 
the 2.16 m telescope at the XingLong Station of the Beijing 
Astronomical Observatory (BAO) in China (Kong \& Cheng 2002).  
Using emission lines of spectra, 74 BCGs with narrow emission lines 
were classified into star-forming galaxies (SFG; Kong et al. 2002).  
The main goal of this project is to measure the current star 
formation rates, stellar components, metallicities, and star 
formation histories and evolution of BCGs; therefore, we are mainly 
interested in these 74 BCGs. The galaxy names were listed in Figure 
1.

\section{Population synthesis results}

\subsection{Data and calculation}

To resolve the stellar components of BCGs, we have used the observed 
equivalent width values of \hbox{Ca\,{\sc ii}}K3933, 
\hbox{H$\delta$}4102, CN4200, G band4301, \hbox{H$\gamma$}4340, 
\hbox{Mg\,{\sc i}+Mg\,{\sc h}}5176 absorption features, the 
continuum fluxes (normalized at 5870\AA) at 3660, 4020, 4510, 6630, 
and 7520{\AA} (Kong et al. 2002) and the empirical population 
synthesis method by Cid Fernandes et al. (2001), which is based on 
spectral group templates built from star clusters, and on Bayes 
theorem and the Metropolis algorithm.  The output is the expected 
values of the fractional contribution (${\bf X}_i$) of each stellar 
component to 
the total flux of galaxy at a reference normalization wavelength, 
such as 5870{\AA}.

\subsection{Stellar population}
Age-grouped results of stellar population synthesis are plotted in
Figure 1. 
${\bf X}_{\rm OLD}$, made up from the sum of all base components
with age $= 10$ Gyr, 
${\bf X}_{\rm INT}$, corresponding to the $10^9$ yr 
intermediate-age bin, 
${\bf X}_{\rm YBC}$, standing for the contribution from the young 
blue stellar populations ($10^7$ -- $10^8$ yr), and ${\bf X}_{\rm 
HII}$ containing the contribution of the power-law component.

\begin{figure}
\psfig{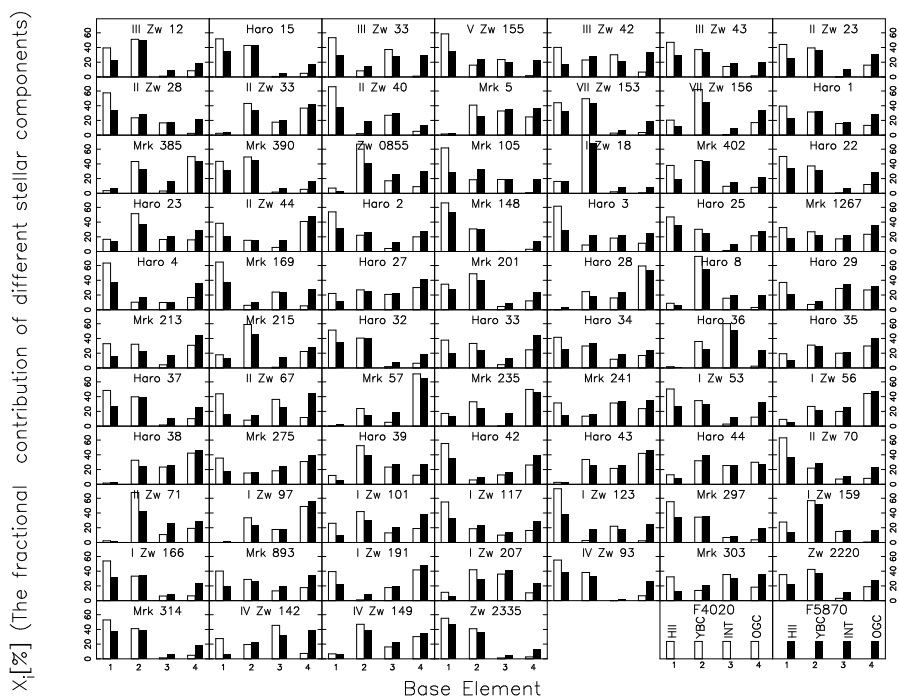}
\caption{Age-grouped synthetic population vector for the 74 BCGs.
The number in the horizontal axis represents different age-grouped 
stellar population.  The vertical axis shows the percentage 
contributions of these age-grouped populations.  Empty histograms 
represent the flux fractions at 4020\AA; and filled histograms 
correspond to the flux fractions at 5870\AA\ respectively.}
\label{fig1}
\end{figure}

A first noticeable result in Figure 1 is that all BCGs show an 
underlying old stellar population.  The presence of large fractions 
of old components indicates that the star formation happened already 
at an early stage, and at a high rate.  It suggests that BCGs are 
old galaxies.  The stellar populations of BCGs present a variety 
of characteristics; the dominant stellar population is different 
in different galaxies. Some BCGs have many young stellar populations. 
However, in others, an intermediate age stellar population makes 
an appreciable contribution. 

Based on the stellar populations, we found BCGs, while sharing some 
common global properties, in fact exhibit a great diversity in the 
star formation histories (SFH). The SFH of BCGs is more complex than 
we thought; we cannot use unification SFH to all BCGs. 

\subsection{Synthesized spectra} 
\begin{figure}
\begin{center}
\psfig{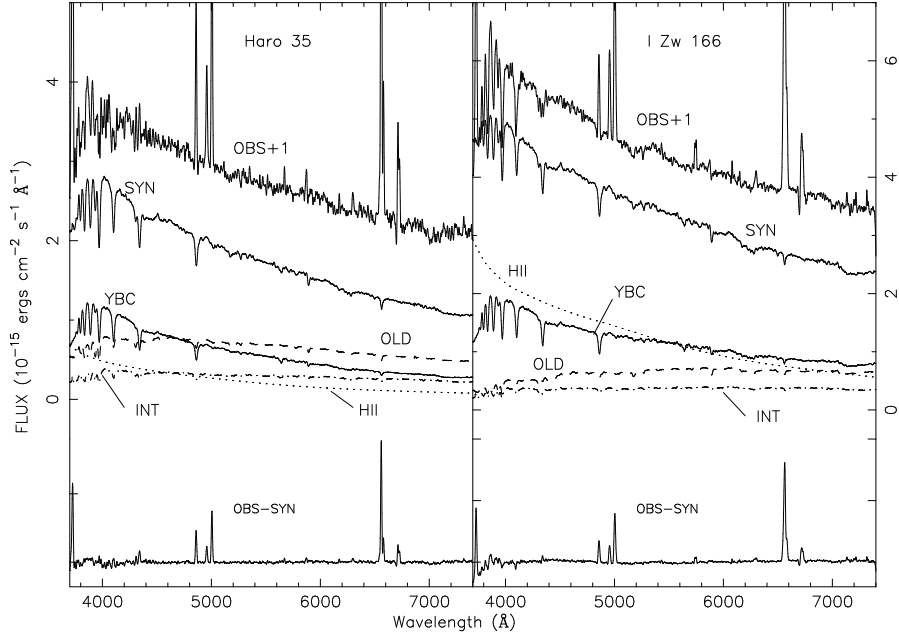}
\end{center}
\caption{Observed spectra of two BCGs, Haro 35 and I Zw 166 
(corrected for reddening; solid line) superposed to the synthesized 
one (dotted line).  Components of the synthesis grouped into old 
(OLD), intermediate age (INT), young (YBC), and HII (HII), shown 
to scale according to their flux fraction contributions at 5870 \AA.  
The emission line spectrum appears in the (OBS-SYN) difference, in 
the lower part of the figure.}
\label{fig2}
\end{figure}

In Figure~\ref{fig2}, as an example, we plotted the empirical 
population synthesis results for 2 galaxies in our sample.
{\bf OBS} represents the observed spectrum of galaxy; 
{\bf SYN} represents the synthetic spectrum, was constructed using 
the 
star cluster templates and the EPS results,
{\bf OBS-SYN} resulting from subtracting {\bf SYN} from {\bf OBS}.

The figure shows that the synthesized spectrum gives a good fit to 
the observed continuum and absorption lines for each galaxy.  It 
suggested that the main energy sources of BCGs are young hot O, B 
stars, which lead to the formation of HII regions around them.  
Another apparent character in Figure 2 is the absorption wing of 
\hbox{H$\beta$}\, and \hbox{H$\gamma$}\, in observed spectrum was 
fitted very well by the synthetic spectrum. We can use this synthetic 
spectrum to subtract the underlying stellar absorption from 
emission line spectrum.
Therefore, the stellar subtracted spectra should be very useful for 
further investigation of physical conditions and chemical abundance 
of the emission line regions of BCGs and will be used to accurate 
determine the element abundance and star formation rate of BCG.

\end{article}
\end{document}